\documentclass[12pt]{amsart}
\usepackage{amsmath,bm, graphicx,color, slashed}
\usepackage{amssymb,bbm}
\usepackage{slashed}
\usepackage[all]{xy}
\xyoption{import}

\def\<{\langle}
\def\>{\rangle}

\def \I{\mathbbm{1}}
\def\H{ {\mathcal H} }

\def\D{ {\mathcal D} }

\def\P{ {\mathcal P} }
\def\A{ {\mathcal A} }
\def\B{ {\mathcal B} }
 
\def\sys{\widehat{\psi}_\A}

\newcommand{\Tr}{\operatorname{Tr}}

\theoremstyle{definition}

\theoremstyle{remark}

\numberwithin{equation}{section}

\begin{document}
\bibliographystyle{amsplain}
\title[Continuum tensor networks]{The continuum limit of a tensor network: a path integral representation}

\author{Christoph Brockt}
\address{Leibniz Universit\"at Hannover, Institute of Theoretical Physics, Appelstra{\ss}e 2, D-30167 Hannover, Germany}
\author{Jutho Haegeman} 
\address{Vienna Center for Quantum Science and Technology, University of Vienna, Austria}
\author{David Jennings}
\address{Controlled Quantum Dynamics Theory Group, Level 12, EEE, Imperial College London, London SW7 2AZ, United Kingdom}
\author{Tobias J.\ Osborne} 
\address{Leibniz Universit\"at Hannover, Institute of Theoretical Physics, Appelstra{\ss}e 2, D-30167 Hannover, Germany}
\email{tobias.osborne@itp.uni-hannover.de}
\author{Frank Verstraete}
\address{Vienna Center for Quantum Science and Technology, University of Vienna, Austria}

\date{\today}

\begin{abstract}
	We argue that the natural way to generalise a \emph{tensor network} variational class to a continuous quantum system is to use the \emph{Feynman path integral} to implement a \emph{continuous tensor contraction}. This approach is illustrated for the case of a recently introduced class of quantum field states known as \emph{continuous matrix-product states} (cMPS). As an example of the utility of the path-integral representation we argue that the state of a dynamically evolving quantum field admits a natural representation as a cMPS. An argument that all states in Fock space admit a cMPS representation when the number of variational parameters tends to infinity is also provided.
\end{abstract}

\maketitle

\section{Introduction}

It is a remarkable fact that a generic state of a quantum system of moderate dimension is unlikely to be physically realisable. This is because physically accessible states occupy only a tiny submanifold of the full hilbert space $\mathcal{H}$, as follows from a simple counting argument (see e.g.\ \cite{poulin:2011a}). The submanifold of physically accessible states has been christened \textit{the physical corner of Hilbert space}; states in this corner exhibit highly nongeneric features: a quantum state chosen at random from $\H$ is massively entangled and locally featureless \cite{clifton:2000a} while, in stark contrast, the manifold of physical states realized in Nature exhibit short-range correlations, and often obey entropy area laws \cite{eisert:2010b}. 

The fact that physically realisable states are not generic has important practical consequences for the calculation of observable properties. Instead of parametrising a physical state using a complete (exponentially large basis) of hilbert space, entropy area laws suggest that there is a more efficient description where only the physical degrees of freedom participating in short-ranged correlations are used. Such a parametrisation, if found, would allow unprecedented insights when combined with the \emph{variational method}.

A compelling example of an efficient parametrisation of the physical corner is found in the study of one-dimensional strongly interacting quantum lattice systems \cite{sachdev:1999a, giamarchi:2004a, auerbach:1994a}. The archetypal variational class here is the set of \emph{matrix product states} (MPS) \cite{fannes:1992a}. The variational method applied to MPS, synonymous with the \emph{density matrix renormalisation group} \cite{white:1992a, schollwock:2011a, schollwoeck:2005a}, has enjoyed unparalleled success in the study of the equilibrium and nonequilibrium dynamics of quantum lattice systems. 

The MPS class has now been extended far beyond the one-dimensional setting to arbitrary quantum lattice systems where they are now understood as examples of \emph{tensor networks}. The two most notable tensor networks are the \emph{projected entangled-pair states} (PEPS) \cite{verstraete:2004a}  and the \emph{multiscale entanglement renormalisation ansatz} (MERA) \cite{vidal:2006a, vidal:2007a}. Both of these classes have been applied with considerable success to the study of strongly correlated physics in two dimensions and beyond for noncritical and critical systems. Tensor networks such as PEPS and MERA are not yet able to explain all the physics of low-dimensional lattice systems, but already many new insights have been obtained.

While the MPS and MERA classes are naturally defined in the presence of a lattice, they have both \cite{verstraete:2010a, osborne:2010a, haegeman:2010a, haegeman:2011a} been generalised in a natural way to continuous quantum systems. Here a basic challenge is that a discrete collection of degrees of freedom is replaced by a \emph{continuous} collection. This means that, for a tensor network state, a contraction of a discrete collection of tensors needs to be replaced by the ``contraction'' of a now continuous set of tensors. Here we argue that the most natural systematic way to achieve this is to replace the tensor contraction with a \emph{path integral} over some now continuous auxiliary degrees of freedom. Thus the core objective of this paper is to obtain and study such a path-integral representation for the cMPS class. This representation will then be used as the basis for higher-dimensional generalisations in a forthcoming paper.

\section{Background: matrix product states, tensor networks, and path integrals}\label{sec:background}
Here we review the MPS class and sketch some of its properties. Our intent is to make this paper accessible to those with a diversity of backgrounds, so we have tried to provide all the necessary prerequisite material and references needed to follow our argument here. Readers with a familiarity with MPS and the DMRG are invited to skim this section lightly to fix notation.

We begin by recalling that any bipartite pure quantum state $|\psi_{\A \B}\>$ admits a Schmidt decomposition $|\psi_{\A \B}\> = \sum_k \sqrt{\lambda_k} |k\>_\A \otimes |k\>_\B$ for some set of local bases of $\A$ and $\B$. For pure states  $|\psi\>$  of one-dimensional quantum spin systems $\A_1 \A_2 \cdots \A_n$ with local dimension $d$, we may perform a Schmidt decomposition iteratively on the bipartitions $[\A_1, \A_1']$, $[(\A_1\A_2), (\A_1\A_2)'], \cdots$ $[\A_n', \A_n]$  (where $X'$ is the complement of $X$ and $[X, Y]$ denotes the particular bipartite split) to obtain the MPS representation \cite{vidal:2003b}
\begin{equation}\label{eq:mps}
	|\psi\rangle = \sum_{j_1, \ldots, j_n = 0}^{d-1} \langle \omega_L|A^{j_1}A^{j_2} \cdots A^{j_n}|\omega_R\rangle |j_1j_2\cdots j_n\rangle.
\end{equation}
Here $A^{j_k}$, $j_k = 0, 1, \ldots, d-1$, is a collection of $d$ matrices of size $D_{k-1}\times D_{k}$, $\langle\omega_L|$ is a row vector of dimension $D_0$, and $|\omega_R\rangle$ is a column vector of dimension $D_n$. The dimensions $D_k$ are called the \emph{bond dimensions} of the MPS and characterize the degree of entanglement entropy across a cut at site $k$. This construction shows that MPS are an\ \emph{expressive} class, meaning that \emph{any} state may be represented as an MPS with a sufficiently large choice of the $D_k$s (the argument applies to \emph{any} pure state). However, in most implementations we simply assume that the bond dimension is constant and truncate it at some value $D_k=D$, which acts as a \emph{refinement parameter} for this class.

Matrix product state representations (\ref{eq:mps}) possess several remarkable properties. The first, and most important, is that they provide an efficient parametrisation of naturally occurring states \cite{hastings:2007a, osborne:2005d, schollwoeck:2005a, schollwock:2011a}; it is now understood that MPS can very efficiently represent both the ground states of models with a spectral gap and also the non-equilibrium dynamics of any quantum spin chain. The second property is that they possess an entropy area law \cite{eisert:2010b}, meaning that the von Neumann entropy of any contiguous block of spins is bounded above by a constant, i.e., the size of the boundary. Another important property of MPS is a \emph{gauge} degree of freedom, so  they supply an over-complete parametrisation. 

A matrix product state (\ref{eq:mps}) is an example of a quantum state known as a \emph{tensor network state} (TNS). To define a TNS one first associates a finite graph $G = (V, E)$ with the quantum system where the physical degrees of freedom, which are quantum spins of dimension $d$, live on the vertices $V$, and the edges $E$ encode \emph{auxiliary} degrees of freedom. To each vertex $v$ we associate a tensor $A_{\alpha_{e_1}\alpha_{e_2}\cdots \alpha_{e_{d_v}}}^{j_v}$ with $d_v+1$ indices, where $d_v$ is the \emph{degree} of the vertex $v$. Each index $\alpha_{e}$ is associated with a corresponding edge $e \in E$ incident with the vertex $v$ and takes values $1, 2, \ldots, D_{e}$; these are the \emph{auxiliary bond indices}. The index $j_v$ is the \emph{physical index} and takes values $0, 1, \ldots, d-1$. The TNS corresponding to this arrangement of tensors is then given by
\begin{equation}
	|\psi\rangle = \sum_{j_{v_1},j_{v_2},\cdots, j_{v_{|V|}}}\mathcal{C}( A^{j_{v_1}} A^{j_{v_2}} \cdots A^{j_{v_{|V|}}})|j_{v_1} j_{v_2} \cdots j_{v_{|V|}}\rangle,
\end{equation}  
where $\mathcal{C}$ denotes the \emph{contraction} of all the auxiliary indices, i.e., each pair of tensor indices associated with each edge are separately summed. Such TNSs may be represented pictorially where we draw a ``leg'' for each index of each tensor and join contracted indices with lines. Physical indices are drawn as unpaired arrows. For example, the simple tensor network resulting from the multiplication of two matrices $\sum_{\beta} A_{\alpha \beta}B_{\beta \gamma}$, is represented by:
\begin{equation}
	\xymatrix{
	 &*+[F-:<3pt>]\txt{$A$}\ar[l]_{\alpha}\ar@{-}[r]^{\beta} &*+[F-:<3pt>]\txt{${B}$}\ar[r]^{\gamma}&.	}
\end{equation}

In the case of an MPS we associate with each tuple of matrices $A^{j_k}$, regarded as a \emph{three-index tensor} $[A^{j_k}]_{\alpha_{k-1}\alpha_{k}}$, the diagram according to
\begin{equation}
	\xymatrix{
	 & & &\\
	  [A^{j_k}]_{\alpha_{k-1}\alpha_{k}}&\equiv&&*+[F-:<3pt>]\txt{$A$}\ar[r]^{\alpha_{k}}\ar[u]_{j_k}\ar[l]_{\alpha_{k-1}}&, \\
	}
\end{equation}
In the pictorial representation the coefficient of $|j_1j_2\cdots j_n\rangle$ for an MPS is depicted as
\begin{equation}
	\xymatrix{
	 & & & & & & \\
	 \langle\omega_L|A^{j_1}A^{j_2} \cdots A^{j_n}|\omega_R\rangle& = &*+[o][F-]{\omega_L}\ar@{-}[r] &*+[F-:<3pt>]\txt{$A$}\ar@{-}[r]^{\alpha_{1}}\ar[u]_{j_1} &*+[F-:<3pt>]\txt{$A$}\ar[r]^{\alpha_{2}}\ar[u]_{j_2}& \cdots & *+[F-:<3pt>]\txt{$A$}\ar[l]_{\alpha_{n-1}}\ar[u]_{j_n}&*+[o][F-]{\omega_R}\ar@{-}[l]}
\end{equation}

The contraction involved in the definition of a tensor network state may also be expressed in terms of a \emph{path integral}. To do this we define the following discrete ``action''
\begin{equation}
	S [ (\alpha_1, \alpha_2 \ldots, \alpha_{|E|}); (j_1, \ldots, j_{|V|}) ] \equiv \sum_{v\in V} -i\log (A^{j_v}_{\alpha_{e_1}\alpha_{e_2}\cdots \alpha_{e_{d_v}}}),
\end{equation}
With this definition, the TNS is given by 
\begin{equation}\label{eq:dfpi}
	|\psi\rangle = \int \mathcal{D}\boldsymbol{\alpha}\,\mathcal{D}\boldsymbol{j} \, e^{iS[\boldsymbol{\alpha}, \boldsymbol{j}]} |\boldsymbol{j}  \rangle,
\end{equation}
where $\int \mathcal{D}\boldsymbol{\alpha}\mathcal{D}\boldsymbol{j}$ denotes here a discrete \emph{sum over all paths} $\boldsymbol{\alpha} = (\alpha_1, \alpha_2, \ldots, \alpha_{|E|})$ and $(j_1, \ldots, j_{|V|})$ with $\alpha_k \in \{1, \ldots, D\}$ and $j_k \in \{1, 2, \ldots, d\}$. 

Intuitively speaking, the way to take a continuum limit of a TNS is to choose the tensors $A^j$ so that as the spacing between the sites goes to zero the \emph{density} of particles/excitations in the system remains constant. In one dimension, for MPS, this is achieved \cite{verstraete:2010a, osborne:2010a} by choosing
\begin{equation}
	\begin{split}
		A^0 &= \mathbb{I} + \epsilon Q \\
		A^1 &= \epsilon R \\
		&\vdots
	\end{split}
\end{equation}
where $Q$ and $R$ are arbitrary $D\times D$ matrices and $\epsilon$ is the lattice spacing. We'll see in the next section that with this choice of $A^j$s the path integral (\ref{eq:dfpi}) reduces, in the limit $\epsilon \rightarrow 0$, to a standard path integral. In the sequel to this paper, we'll show that a similar recipe works for any sufficiently regular lattice.

\section{Path integrals and continuous matrix product states}\label{sec:picmps}
Continuous matrix product states are a variational class of states for one-dimensional quantum fields. We focus on the bosonic case with field annihilation and creation operators $\sys(x)$ and $\sys^\dagger(x)$ such that $[\sys(x), \sys^\dagger(y)]=\delta(x-y)$. A cMPS is then defined in terms of the quantum field $\mathcal{A}$ and an auxiliary $D$-level quantum system $\mathcal{B}$ by
\begin{equation}\label{eq:cmps2}
\boxed{|\chi \> = \langle \omega_L| \P \exp\left[-i\int_0^l K(s)\otimes\mathbb{I} + iR(s)\otimes \sys^\dag(s) - iR^\dag(s)\otimes \sys (s) \,ds\right]|\omega_R\rangle |\Omega_{\mathcal{A}}\rangle,}
\end{equation}
where $K$ is a $D\times D$ hermitian matrix and $R$ is $D\times D$ complex matrix, $|\omega_{L,R}\>$ are $D$-dimensional states of the auxiliary system $\B$, $\sys (s)$ is a bosonic field operator on the physical system $\A$, $|\Omega_{\mathcal{A}}\rangle$ is the fock vacuum, and $\mathcal{P}$ denotes path ordering.

\subsection{A path integral for the auxiliary system}
We wish to reformulate the cMPS state (\ref{eq:cmps2}) so that expectation values for the auxiliary system are recast as path-integral expressions. The motivation for this is two-fold: firstly, to facilitate the passage to higher-dimensional cMPS states; and secondly, to make manifest the symmetries of the physical state in terms of symmetries of an action for the auxiliary system. Our discussion is centred on the case of a single bosonic field in (1+1) dimensions; the generalisation to spinor and vector fields follows easily, and we only comment on the modifications required.

Write a basis for the hilbert space $\mathcal{H}_{\mathcal{B}}$ of $\mathcal{B}$ as $\{|j\rangle\,|\, j = 0, 1, \ldots, D-1 \}$. We enlarge this space via second quantisation, i.e., by either introducing bosonic annihilation and creation operators $b_j$ and $b_j^\dag$ according to the canonical commutation relations:
\begin{equation}
	[b_j, b_k^\dag] = \delta_{j,k}, \quad j = 0, 1, \ldots, D-1,
\end{equation}
with all other commutators vanishing, or fermionic annihilation and creation operators $c_j$ and $c_j^\dag$ according to the canonical anticommutation relations
\begin{equation}
	\{c_j, c_k^\dag\} = \delta_{j,k}, \quad j = 0, 1, \ldots, D-1,
\end{equation}
with all other anticommutators vanishing. The configuration space for our enlarged auxiliary system is that of \emph{fock space} $\mathfrak{F}_{\pm}(\mathcal{H}_{\mathcal{B}})$, where the $\pm$ subscript indicates either bosonic or fermionic fock space.

The connection between $\mathcal{H}_{\mathcal{B}}$ and our enlarged fock space $\mathfrak{F}_{\pm}(\mathcal{H}_{\mathcal{B}})$ is made, in the bosonic case, by identifying $\mathcal{H}_{\mathcal{B}}$ with the single-particle sector via
\begin{equation}
	|j\rangle_\mathcal{B} \equiv b_j^\dag |\Omega\rangle_{\mathcal{B}},
\end{equation}
or, in the fermionic case,
\begin{equation}
	|j\rangle_\mathcal{B} \equiv c_j^\dag |\Omega\rangle_{\mathcal{B}},
\end{equation}
where $|\Omega\rangle_{\mathcal{B}}$ is the fock vacuum. We identify, whenever clear from the context, states $|\omega\rangle \in \mathcal{H}_{\mathcal{B}}$ with their single-particle counterparts in $\mathfrak{F}_{\pm}(\mathcal{H}_{\mathcal{B}})$.

Using this embedding, a cMPS (\ref{eq:cmps2}) is equivalent, in the bosonic case, to
\begin{equation}
|\chi\> = \<\omega_L | U(l,0)|\omega_R\> |\Omega_\A\> = \<\omega_L | \P \exp \left[-i\int_0^l F(s)\, ds \right]|\omega_R\> |\Omega_\A\>,
\end{equation}
where $F$ is a one-parameter set of field operators on $\A \B$, generated by $U(l,0)$ and given is by
\begin{eqnarray}
F(s) =  \sum_{j,k=1}^D K^{jk}(s) b_j^{\dag} b_k  \otimes\I + iR^{jk}(s)  b_j^{\dag}b_k\otimes \sys^\dag(s) -iR^{*\, kj}(s)  b_j^{\dag}b_k\otimes \sys^\dag(s),
\end{eqnarray}	
This equivalence of definitions follows from the fact that $F(s)$ is particle-number conserving on system $\B$ (i.e. its action on $\B$ is through terms of the form $b_j^\dagger b_k$ only), and so we remain in the single-particle sector throughout. The fermionic version is identical except that $b_j$ operators are replaced with $c_j$s. 

The parameter $s$ can be regarded as a \emph{time coordinate} for the auxiliary system. We then obtain a path integral by dividing $[0,l]$ into small intervals $s_0=0, s_1, s_2, \dots s_N=l$ with uniform spacing $s_{k+1}-s_k = \epsilon$, so that $U(l,0) = U(l,l-\epsilon)U(l-\epsilon,l-2\epsilon)\cdots U(\epsilon,0)$, and insert resolutions of the identity between each term. Our choice of resolution is, in the bosonic case, in terms of coherent states of the auxiliary system, defined as $|\phi_k\> = \exp[ \phi_k b_k^\dagger - \phi^*_k b_k]|\Omega_\B\>$:
\begin{equation}\label{eq:boscohres}
\I = \frac{1}{\pi^N} \int \prod_k d^2\phi_k \,|{}\otimes_k \phi_k\rangle\<\otimes_k \phi_k|,
\end{equation}
where $N = l/\epsilon$.
In the fermionic case we exploit fermion coherent states of the form $|\phi_k\> = \exp[ c_k^\dagger\phi_k  - \phi^*_kc_k]|\Omega_\B\>$, where $\phi_k$ are now \emph{Grassmann-valued}. Apart from the use of anticommuting Grassmann numbers the fermionic calculation mirrors the bosonic case in essentially all other respects; we thus focus on the details of the bosonic calculation and write out the fermionic case at the end.

After the resolution (\ref{eq:boscohres}) has been placed between each term we end up with a product of transition amplitudes of the form $\<\otimes_k \phi_k (s+\epsilon)|U(s+\epsilon ,s)|\otimes_{k'} \phi_{k'} (s)\> \approx \<\otimes_k \phi_k(s+\epsilon)| \I -i\epsilon F(s)|\otimes_{k'}\phi_{k'} (s)\>$. We then make use of the expression
\begin{equation}\label{coherentoverlap}
\<\otimes_k \phi_k(s+\epsilon)|\otimes_{k'} \phi_{k'} (s)\> = \exp \left[-\frac{1}{2} \sum_{k=1}^D  |\phi_k(s+\epsilon)|^2 + |\phi(s)|^2 - 2 \phi_k^*(s+\epsilon)\phi_k(s) \right]
\end{equation}
and the assumption that only smooth variations of $\phi_k(s)$ contribute, which allows us to expand the terms in the exponential and obtain, in the continuum limit $\epsilon \rightarrow 0$,
\begin{eqnarray}
|\chi\> &=& \int \prod_k \D^2 \phi_k\, \P \exp \left[ i \widehat{S} (\phi_k, \phi_k^*) \right] |\Omega_\A\>,
\end{eqnarray}
where the path integral is over $D$ complex fields and $\widehat{S}$ is an \emph{operator-valued} action given by
\begin{equation}
	\widehat{S} = \int ds \left(i \phi^\dagger \partial_s \phi - \phi^\dagger K\phi - i(\phi^\dagger R\phi)\sys^\dagger + i(\phi^\dagger R^\dag\phi)\sys^\dagger\right),
\end{equation}  
where we've written $\{ \phi_k\}$ as a vector $\phi$. However, since the field operator $\sys^\dagger(s)$ commutes with $\sys(s')$ and $\sys^\dag(s')$ at all other points $s'$ the ordering over the auxiliary time variable is trivial and we can simply write the path integral as
\begin{equation}\label{pathcmps}
\boxed{|\chi\> = \int \D^2\phi\, \exp \left[ iS (\phi, \phi^\dagger) \right] |\Phi\>}
\end{equation}
where $|\Phi\rangle$ is a \emph{physical} field coherent state 
\begin{equation}
	|\Phi\> \equiv \exp \left[\int \Phi(s) \sys^\dagger(s) - \Phi^*(s) \sys (s) \, ds \right] |\Omega_\A\>,
\end{equation}
$\Phi(s) = \phi^\dagger R\phi$, and the complex action $S$ is given by
\begin{equation}
	S(\phi, \phi^\dagger) = \int ds \left(i \phi^\dagger \partial_s \phi - \phi^\dagger K\phi\right).
\end{equation}
This formulation (\ref{pathcmps}) of the one-dimensional cMPS state as a path integral is one of our guiding expressions for the generalisation to higher-dimensional scenarios which we describe in a sequel paper. The fermionic case is identical, except that $\phi$ is now a vector of Grassmann fields. Of course, both the bosonic and fermionic cases yield exactly the same physical state $|\chi\rangle$, since they coincide on the 1-particle sector. Notice that we've dropped the limits from the integrals; the expression (\ref{pathcmps}) makes equal sense for quantum systems on $[0,l]$ as for the infinite case $(-\infty, \infty)$.

\subsection{Interpretation of the cMPS path integral}

Let's pause to interpret (\ref{eq:cmps2}) for a moment. What these equations are saying is that a cMPS is a superposition of coherent states $|\Phi\rangle$ with some weighting $e^{iS}$. The standard intuition concerning coherent states is that they are the ``most classical'' states of a quantum system due to their saturation of the Heisenberg uncertainty relation. Thus, (\ref{pathcmps}) tells us that a cMPS is a superposition of ``classical'' field states centred around classical field configurations $\Phi:\mathbb{R}\rightarrow \mathbb{C}$ in phase space. These field configurations $\Phi$ themselves are scalar functions of a vector of \emph{auxiliary} classical fields $\phi:\mathbb{R}\rightarrow \mathbb{C}$. By interpreting that spatial variable $s$ as a temporal variable one can understand the action $S$ for these auxiliary fields as that of a $(0+1)$-dimensional quantum field, i.e., ordinary quantum mechanics. 

One has the picture of an auxiliary system undergoing a classical trajectory of its discrete variables, however to gain information (by measurement) about a dynamically evolving quantum system we inevitable disturb it because of the back-action of the quantum measurement. The closest representation of the dynamics in this quantum setting is to \emph{continuously monitor} the evolving auxiliary system with a sequence of infinitesimally weak measurements \cite{caves:1987a}. By exploiting von Neumann's prescription for quantum measurement this process is then understood as \emph{entangling} the auxiliary system and an infinite collection of \emph{meter} systems. The combined auxiliary system+meter collection undergoes completely positive dynamics. In the continuum limit the meter systems constitute a quantum field with one extra spatial dimension, the reduced state of the meters alone is a quantum state. The cMPS coherent field state is then an imprint of the discrete trajectory, and is as classical a record as possible. The stength and manner of this imprint is entirely contained in the particular coupling $R(t)$. Each trajectory for the auxiliary system contributes a coherent field state, and the cMPS is simply a superposition of ``classical'' trajectories with the according weighting by the action $S$. 

\begin{figure}
\includegraphics{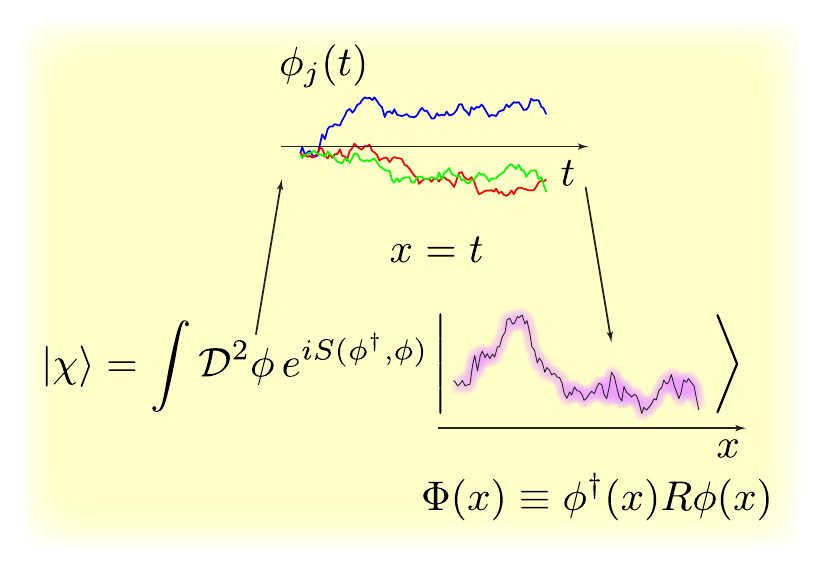}
\caption{An illustration of the coherent-state path integral representation for a cMPS state $|\chi\>$: here a sample classical trajectory for the (in this case, three) auxiliary fields is depicted above. These classical trajectories are then combined via $\phi^\dag(x)R\phi(x)$ into a single complex scalar trajectory $\Phi(x)$. The field coherent state is then represented diagrammatically via the ket with the purple smeared trajectory (the idea is that a field coherent state is like a smeared-out classical configuration centred on the classical configuration $\varphi(x) \propto \text{Im}(\Phi(x))$). The formula for the resulting cMPS is then a superposition of such coherent states weighted by the ``action'' $S$.}
\end{figure}

\section{Completeness of the cMPS class}
In this section we show that an arbitrary one-dimensional quantum field state is expressible as a cMPS, as long as the bond dimension $D$ is allowed to become arbitrarily large. The argument we present here is for the case of bosonic fock space $\mathfrak{F}_-(L^2([0,l]))$ on a \emph{finite} interval $[0,l]$, however, it is expected to be valid in the more general case of the interval $(-\infty, \infty)$. 

The argument is rather simple and relies on three facts. The first is that an arbitrary quantum field coherent state 
\begin{equation}
	|\Phi\> \equiv \exp \left[\int_0^l \Phi(s) \sys^\dagger(s) - \Phi^*(s) \sys (s) \, ds \right] |\Omega_\A\>,
\end{equation}
is exactly representible as a cMPS $|\chi(\Phi)\rangle$ with bond dimension $D=1$. This follows upon taking $K$ and $R$ to be the one-dimensional matrices $K(s) = 0$ and $R(s) = \Phi(s)$. The boundary vectors $|\omega_L\rangle$ and $|\omega_R\rangle$ are simply taken to be equal to $1$. The next fact we require is that the set of all field coherent states is \emph{dense} in Fock space, meaning that an arbitrary field state $|\Psi\rangle \in \mathfrak{F}_-(L^2([0,l]))$ may be approximated arbitrarily well by an increasing linear combination of field coherent states:
\begin{equation}
	\sum_{l=0}^N c_j|\Phi_j\rangle \xrightarrow{N\rightarrow \infty}|\Psi\rangle.
\end{equation}
The final fact we need is that a linear combination $|\chi\rangle = c_1|\chi_1\rangle + c_2|\chi_2\rangle$ of two cMPS $|\chi_1\rangle$ and $|\chi_2\rangle$ with bond dimensions $D_1$ and $D_2$, respectively, is again a cMPS with bond dimension $D=D_1+D_2$ and parameters $K = K_1\oplus K_2$, $R = R_1\oplus R_2$, $\langle \omega_L| = (c_1 \langle\omega_{L,1}|\oplus c_2\langle \omega_{L,2}|)$, and $|\omega_R\rangle = |\omega_{R,1}\rangle \oplus |\omega_{R,2}\rangle$.

Putting these facts together allows us to deduce that 
\begin{equation}
	|\chi_N\rangle \equiv \sum_{l=0}^N c_j|\Phi_j\rangle
\end{equation}
is a sequence of cMPS with bond dimensions $D_N = N$ that tend, in the limit, to an arbitrary state $|\Psi\rangle$ in fock space. Thus we have confirmed the \emph{completeness} or \emph{expressiveness} property of the cMPS variational class in one dimension. It is worth noting that the argument we present here is by no means the most economical: there are, exploiting gauge invariance, almost certainly more efficient sequences of representations tending to the state $|\Psi\rangle$ using lower bond dimensions. Indeed, as we argue in the next section, a more economical representation of a physical field state is strongly suggested by the path integral representation.

It is worth noting that in the previous subsection we showed that an arbitrary cMPS is a superposition of field coherent states. Here we've shown the converse: an arbitrary superposition of field coherent states is also a cMPS.

\section{cMPS representation of physical quantum field states via path integrals}

In this section we show how naturally occurring physical states are efficiently represented via cMPS. Intuitively, the argument presented here is a continuum version of that appearing in \cite{banuls:2009a}, i.e., we exchange the role of space and time and contract the (continuum) tensor network for the state of a dynamically evolving field along the spatial axis first, regarding it is a temporal axis.

We frame our results in the bosonic setting, although the extension to the fermion case is straightforward. The generic situation we consider is therefore that of bosons moving in $\mathbb{R}$ with field annihilation $\widehat{\psi}(x)$ and creation $\widehat{\psi}^\dag(x)$ operators obeying the canonical commutation relations
$	[\widehat{\psi}(x), \widehat{\psi}^\dag(y)] = \delta(x-y), $
with all other commutators vanishing. 

The hamiltonian, in second quantisation, is taken to have the standard kinetic+potential energy form:
\begin{equation}
	\widehat{H} = \widehat{T} + \widehat{V} + \widehat{W}, 
\end{equation}
where
\begin{equation}
	\widehat{T} = \int \frac{d\widehat{\psi}^\dag(x)}{dx}\frac{d\widehat{\psi}(x)}{dx}\, dx, \hspace{0.3cm}
	\widehat{V} = \int V(x)\widehat{\psi}^\dag(x)\widehat{\psi}(x) \, dx,
\end{equation}
and with an interaction potential
\begin{equation}
	\widehat{W} = \int W(x-y) \widehat{\psi}^\dag(y)\widehat{\psi}^\dag(x)\widehat{\psi}(x) \widehat{\psi}(y) \, dxdy.
\end{equation}
For simplicity we will concentrate here on ultralocal interactions, i.e. $W(x-y) = w\delta(x-y)$, with $w$ being a constant.

We consider a field system, initialized in the physical state $|\varphi (0) \> \in \H_\A$ and allow it to evolve under the full hamiltonian $\widehat{H}$ for a time $T$, until it reaches the state $|\varphi (T)\> = e^{-iT\widehat{H}} |\varphi (0)\>$. Our task is to show that the final state $|\varphi (T)\>$ admits a natural description in terms of the cMPS path integral representation, which can be interpreted instead as a (virtual) process in which some additional auxiliary system $\H_\B$ undergoes dissipative dynamics that couple it to the physical field system and on completion generates $|\varphi(T)\>$. To avoid confusion, in the case of the \emph{physical} evolution of the field system $\H_\A$ we use $x$ for the spatial coordinate, and $t$ for the physical time coordinate, while for the virtual process in which the auxiliary system $\H_\B$ couples to the physical field $\H_\A$ we use $s$ for the virtual time coordinate of $\H_\B$ and label \emph{subsystems} of $\H_\B$ with the parameter $\beta$. The construction that follows will roughly amount to reinterpreting the field variables $(x,t)$ as $(s,\beta)$ within a quite physically distinct setting.

Our first move is to reformulate the physical field evolution in terms of a path integral expression over coherent states. The construction proceeds, as usual, by discretising the time interval $[0, T]$ into $n$ pieces of length $\epsilon = T/n$ and writing
\begin{equation}
	|\varphi(T)\rangle = (e^{-i\epsilon\widehat{H}})^n |\varphi(0)\rangle.
\end{equation} 
We suppose, for simplicity, that the initial state $|\varphi(0)\rangle$ is a field coherent state.

As in the construction of the auxiliary action, we insert a resolution of the identity, in terms of 1-d field coherent states $|\Phi(t)\> :=\exp [\int \! dx  \,\Phi(x,t) \widehat{\psi}^\dagger(x) - \Phi^*(x,t) \widehat{\psi} (x)] |\Omega\>$, between each application of $e^{-i\epsilon\widehat{H}}$. 
Expanding up to first order, and using the key overlap equation (\ref{coherentoverlap}) we find that an infinitesimal advance for the physical system is described by

\begin{equation}
	\langle \Phi(t)|  e^{-i\epsilon\widehat{H}} |\Phi(t-\epsilon) \rangle \approx e^{-\frac{\epsilon}{2}\int {\Phi^*(x,t)}\partial_t\Phi(x,t)  - \partial_t{\Phi^*(x,t)}\Phi(x,t)     -i\epsilon \H(\Phi^*(x,t), \Phi(x,t))\, dx},
\end{equation}
with a hamiltonian density $\H(x,t)$ given by
\begin{equation}
	\H(\Phi^*(x,t), \Phi(x,t )) = |\partial_x \Phi (x,t)|^2 + \ V(x)|\Phi (x,t)|^2+  w |\Phi (x,t)|^4  
\end{equation}
Summing over each time interval yields the expression
\begin{equation}\label{eq:realtime}
	|\varphi(T)\rangle = \int \mathcal{D}\Phi \mathcal{D}\Phi^* e^{iS(\Phi, \Phi^*)}|\Phi(T)\rangle,
\end{equation}
being a superposition of physical coherent states described by $\Phi(x,T)$ at time $t=T$, and with the action
\begin{equation}
	S(\Phi, \Phi^*) = \int_0^T \int_{-\infty}^\infty  i{\Phi^*(x,t)}\partial_t\Phi(x,t) - \H(\Phi^*(x,t), \Phi(x,t)) \, dxdt.
\end{equation}
The lower limit of this path integral is $\Phi (x,0) = \varphi(x,0)$ while the upper limit is unconstrained.

The path integral form of $|\varphi (T)\>$ is suggestive of how an auxiliary system should couple to the physical system in order to generate $|\varphi (T)\>$ under (virtual) dissipative dynamics. Since we wish the auxilary system $\H_\B$ to sweep over the length of the physical field the time parameter for the process $s$, should correspond to the physical spatial variable $x$.

To capture this idea we could subdivide the auxiliary system $\H_\B$ into harmonic oscillator subsystems as $\H_\B = \otimes_\beta \H_\beta$, labelled by some variable $\beta$, however since $t$ is a continuous variable we shall effectively be taking the limit in which $\H_\B$ is an \emph{auxiliary} complex field where $\beta$ is its spatial coordinate and the auxiliary system has \emph{spatial} extent from $\beta = 0$ to $ \beta =T$. The key point is that spatial couplings (along $\beta$) within the hamiltonian of the auxiliary system can be used to simulate the physical dynamics that generates $|\varphi (T)\>$, as the auxiliary system sweeps out over the physical field, and couples to it through a natural interaction term.

For the auxiliary variables we use $\widehat{z}_0(s,\beta)$ and $\widehat{z}_1(s,\beta)$, which we can combine into a single complex field as $\widehat{z}= \widehat{z}_0 + i\widehat{z}_1$. The free hamiltonian of the auxiliary system is taken to be
\begin{multline}\label{freeham}
	\widehat{K} (s) = \int_0^T -\frac{1}{4}\widehat{p}_0(s,\beta)^2 - \frac{1}{4}\widehat{p}_1(s,\beta)^2 + V(\widehat{z}_0(s,\beta)^2+\widehat{z}_1(s,\beta)^2)  + w(\widehat{z}_0(s,\beta)^2+\widehat{z}_1(s,\beta)^2)^2 \\
- i(\widehat{z}_0(s,\beta)-i\widehat{z}_1(s,\beta))\partial_\beta (\widehat{z}_0(s,\beta)+i\widehat{z}_1(s,\beta))\, d\beta,
\end{multline}
and where $\widehat{p}_0$ and $\widehat{p}_1$ are the momenta conjugate to $\widehat{z}_0$ and $\widehat{z}_1$.

The form of (\ref{eq:realtime}) suggests that the interaction term coupling the auxiliary and physical systems be taken to be the continuous measurement interaction in which the physical system $\H_\A$ is interpreted as continuously measuring the `observable' $\widehat{z} = \widehat{z}_0 + i\widehat{z}_1$. This is obtained as the continuum limit of the coupling 
\begin{equation}\label{interaction}
	\widehat{H}_{\mathrm{int}}(s) =  i\epsilon\sum_{j\in \mathbb{Z}} \delta(s-j\epsilon) \left[ \widehat{z}(s,\beta=T)\otimes \widehat{\psi}^\dag_{j_\A} - \widehat{z}^\dagger(s,\beta=T)\otimes \widehat{\psi}_{j_\A} \right],
\end{equation}
in other words, the physical system only couples to the extreme edge of the auxiliary system at the (auxiliary) spatial point $\beta=T$. Here $\widehat{\psi}_{j_\A} \equiv \frac{a_{j_\A}}{\sqrt{\epsilon}},$ and $a_{j_\A}$ is the operator which annihilates a boson with wavefunction $\frac{1}{\sqrt{\epsilon}}\chi_{[(j-1)\epsilon, j\epsilon)}(x)$ for the physical system.

It is now a case of checking that the composite system $\H_\A\otimes \H_\B$, evolving under the full hamiltonian $\widehat{H}_{\mathrm{tot}} = \widehat{K}+\widehat{H}_{\mathrm{int}}$ for auxiliary time from $s=-\infty $ to $s=+\infty$ will indeed generate the desired field state $|\varphi (T)\>$ as expressed in the path integral form (\ref{eq:realtime}). The calculation proceeds in a similar manner to the earlier cMPS path integral calculation evolving under the composite hamiltonian $\widehat{H}_{\mathrm{tot}}$, however for our resolution of the identity at auxiliary time $s$ we use the complete set of states $\{|z(s)\>\}$ given by
\begin{eqnarray}
|z(s)\> = |z_0(s,0), z_0(s,\epsilon), \cdots z_0 (s, T) ; z_1(s,\epsilon), z_1(s,2\epsilon), \cdots z_1 (s, T) \>,
\end{eqnarray}
which we express in the discretized setting with oscillators located at $\beta=0,\epsilon, 2\epsilon,\dots, T$. A straightfoward calculation gives that
\begin{equation}
\<\omega_L|\mathcal{P}e^{-i\int{\widehat{H}_{\mathrm{tot}}(s)ds}}|\omega_R\> |\Omega\>
= \int \mathcal{D}^2z(s,\beta) \mathcal{D}^2p(s,\beta)\ \exp{[iS'(p_0,p_1,z_0,z_1)]} |\Omega_\A\>
\end{equation}
where we have the action
\begin{equation}
S' = \int_{-\infty}^\infty ds\int_{0}^Td\beta  \, (p_0\dot{z}_0 + p_1\dot{z}_1 - K(p,z) - iz(s,\beta=T)\widehat{\psi}^{\dagger}_\A + iz^*(s,\beta=T)\widehat{\psi}_\A).
\end{equation}
Consequently, by identifying $z(s, \beta=T)$ with $\Phi(x,T)$ we see that the evolved physical state $|\varphi(T)\>$ can be represented by a cMPS with free hamiltonian $K$ given by (\ref{freeham}) and interaction given by (\ref{interaction}).

The cMPS representation that we have constructed involves an infinite dimensional auxiliary system where integration over $\beta$ corresponds to a continuum summation over the auxiliary indices; this is not unexpected since the auxiliary system faithfully simulates the entire dynamical history of the physical field. However, the local character of the interaction term implies that we can obtain $|\varphi (T)\>$ equally well from the coupling of a single auxiliary oscillator to the physical field, with the composite system now undergoing a more general completely-positive map (instead of a unitary interaction). Specifically, the above calculation has shown that $|\varphi (T)\> = \< \omega_L | U |\omega_R\> |\Omega\>$, or more generally $|\varphi (T)\>\< \varphi (T)| = \Tr_{\mathrm{aux}} [ U (\omega \otimes |\Omega\>\<\Omega| )U^\dagger] $ for some operator $U$ on the joint system and auxiliary state $\omega$, but which can now be written as $\Tr _{\beta=T}[\, \Tr_{\beta \ne T} [U (\omega \otimes |\Omega\>\<\Omega| )U^\dagger]]  = \Tr_{\beta=T} [ \mathcal{E} (\omega_\beta \otimes |\Omega\>\<\Omega| )]$ for some completely-positive map $\mathcal{E}$ defined on the physical field and oscillator at $\beta=T$. By truncation of the oscillator hilbert space, and simulation of the evolution $\mathcal{E}$ we may thus obtain an efficient cMPS description of $|\varphi(T)\>$ in terms of a purely discrete auxiliary quantum system.

\section{Conclusions}

In this paper we have investigated the cMPS variational class of quantum field states from a path integral perspective. Building on the the observation that discrete MPS representations can be viewed as path integral sums that couple discrete auxiliary and physical `trajectories', we have constructed a natural path integral representation of cMPS as a sum over coherent field states. This representation is physically appealing, allows us to show the completeness of the cMPS class, and guides the construction of higher dimensional continuum limit of tensor network states. The construction also allows a natural representation of a dynamically evolved physical field state.

\section*{acknowledgements} 
Helpful discussions with Henri Verschelde are gratefully acknowledged. This work was supported by  EU grants QUERG and QFTCMPS,  FWF SFB grants FoQuS and ViCoM, and by the cluster of excellence EXC 201 Quantum Engineering and Space-Time Research. D.\ J.\ is supported by the Royal Commission for the Exhibition of 1851. 

\providecommand{\bysame}{\leavevmode\hbox to3em{\hrulefill}\thinspace}
\providecommand{\MR}{\relax\ifhmode\unskip\space\fi MR }
\providecommand{\MRhref}[2]{%
  \href{http://www.ams.org/mathscinet-getitem?mr=#1}{#2}
}
\providecommand{\href}[2]{#2}

\end{document}